\documentclass[letterpaper, 10 pt, conference]{ieeeconf}  % Comment this line out if you need a4paper

\IEEEoverridecommandlockouts                              % This command is only needed if 
\usepackage{amssymb}  % assumes amsmath package installed
\usepackage{multirow}

\usepackage{marvosym}
\usepackage{makecell}

\usepackage{graphics} % for pdf, bitmapped graphics files
\usepackage{epsfig} % for postscript graphics files
\usepackage{times} % assumes new font selection scheme installed
\usepackage{stfloats}
\usepackage{enumerate}

\title{\LARGE \bf
Can Quadruped Guide Robots be Used as Guide Dogs?
}

\author{Luyao Wang$^{1}$*, Qihe Chen$^{1}$*, Yan Zhang$^{1}$, Ziang Li$^{1}$, \\ Tingmin Yan$^{1}$, Fan Wang$^{1}$, Guyue Zhou$^{1}$ and Jiangtao Gong$^{1}$\textsuperscript{\Letter}% <-this % stops a space
\thanks{*These authors contributed equally to this work.}
\thanks{$^{1}$Institute for AI Industry Research (AIR), Tsinghua University, 10080, Haidian District, Beijing, P.R.China.
        {\tt\small secondnamefirstname@air.tsinghua.edu.cn}}%
}

\begin{document}

\maketitle
\thispagestyle{empty}
\pagestyle{empty}

%%%%%%%%%%%%%%%%%%%%%%%%%%%%%%%%%%%%%%%%%%%%%%%%%%%%%%%%%%%%%%%%%%%%%%%%%%%%%%%%
\begin{abstract}
Quadruped robots have the potential to guide blind and low vision (BLV) people due to their highly flexible locomotion and emotional value provided by their bionic forms.
However, the development of quadruped guide robots rarely involves BLV users' participatory designs and evaluations.
In this paper, we conducted two empirical experiments both in indoor controlled and outdoor field scenarios, exploring the benefits and drawbacks of quadruped guide robots.
The results show that the nowadays commercial quadruped robots exposed significant disadvantages in usability and trust compared with wheeled robots.
It is concluded that the moving gait and walking noise of quadruped robots would limit the guiding effectiveness to a certain extent, and the empathetic effect of its bionic form for BLV users could not be fully reflected.
Based on the findings of wheeled robots and quadruped robots' advantages, we discuss the design implications for the future guide robot design for BLV users. 
This paper reports the first empirical experiment about quadruped guide robots with BLV users and preliminary explores their potential improvement space in substituting guide dogs, which can inspire the further specialized design of quadruped guide robots.

\end{abstract}
\graphicspath{{pic/}}

%%%%%%%%%%%%%%%%%%%%%%%%%%%%%%%%%%%%%%%%%%%%%%%%%%%%%%%%%%%%%%%%%%%%%%%%%%%%%%%%
\section{INTRODUCTION}
Statistics show from the World Health Organization that around 2.2 billion people worldwide suffer from visual impairment \cite{kuriakose2022tools}. Facing uncertain and complex road conditions, safe and efficient travel has become one of the biggest challenges for the BLV groups \cite{giudice2008blind}. 
A well-trained guide dog is an optimal solution. Guide dog users reported enhanced travel independence and comfort \cite{lloyd2008guide1, lloyd2008guide2}. 
Other than door-to-door accurate navigation, guide dogs have advantageous biological attributes compared with other guiding methods, such as providing more sense of trust and confidence~\cite{hauser2014understanding, whitmarsh2005benefits}. 
%Moreover, the rigid connection structure of the harness can offer grounded kinesthetic feedback\cite{hauser2014understanding, whitmarsh2005benefits}. 
However, facing the needs of the BLV community, the cost of training and caring for guide dogs is so high \cite{bekoff2019unleashing} that many BLV people have no access to these resources, especially in developing countries. 

\begin{figure}
    \centering
    \includegraphics[width=7.5cm]{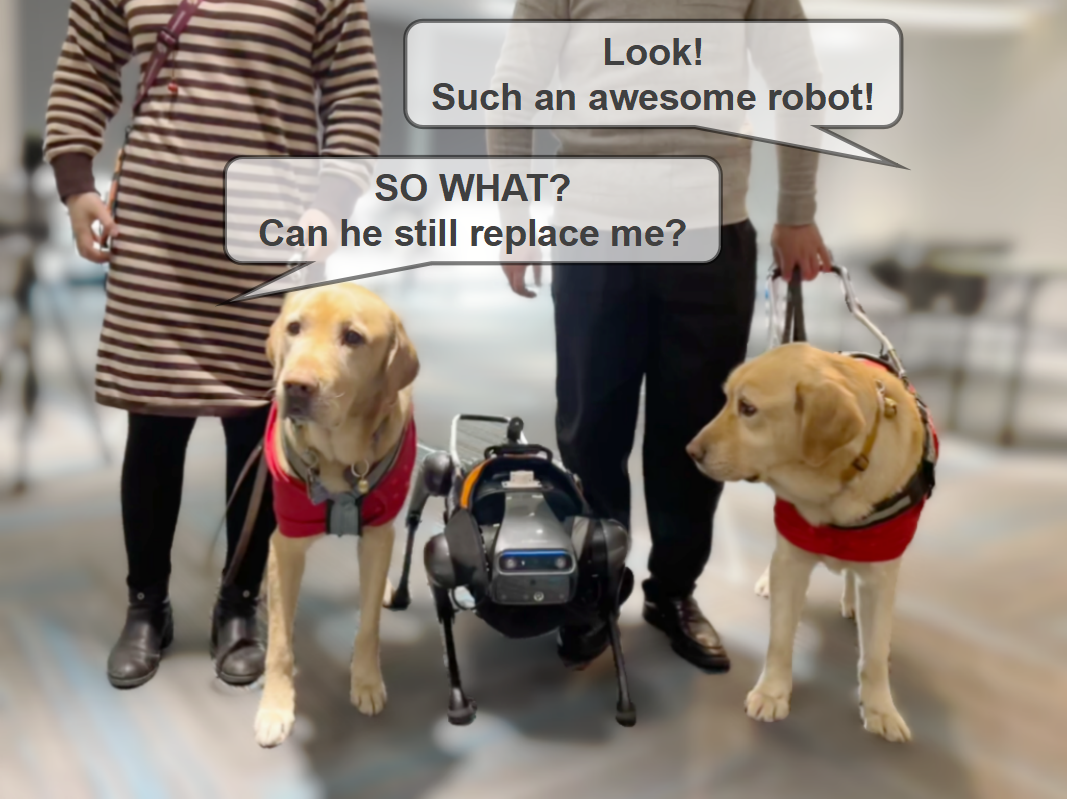}
    \caption{Can Quadruped Guide Robots be Used as Guide Dogs?}
    \label{fig:main_figure}
\end{figure}

Due to the similarity with guide dogs in form and functionality, quadruped robots are expected to compensate for the disadvantages of guide dogs and still exert their advantages. 
Therefore, the implementation in guiding has attracted extensive attention from academia and industry~\cite{chen2021fuzzy,xiao2021robotic,wang2021navdog}, but lack of empirical experiments to verify the feasibility with BLV users.
When discussing functional advantages of guiding, the bionic quadruped robot is an essential grounded robot form \cite{tobita2018structure, guerreiro2019cabot, capi2012development}, and famous for biologically inspired dynamic gait movements and self-stability \cite{li2011research} to have strong flexibility and excellent adaptability even in outside. 
%Regarding bionic guide robots, preliminary research has been conducted \cite{chen2021fuzzy, xiao2021robotic}. 
In \cite{chen2021fuzzy}, a quadruped robot based on the fuzzy control method can maintain stable performance in various road surface changes. A hybrid physical human-robot interaction model incorporating a guide leash was introduced in \cite{xiao2021robotic}, and the ability of the leash to slack and tighten allowed the robot to guide users through narrow spaces. However, there is still a lack of human-centered exploration of robot design and control adjustments to adapt to the preference and utilization of BLV users.%However, these attempts do not show that quadruped robots have functional and emotional advantages in guiding. 

% Bionic properties of robots refer to imitating, replicating, or recreating the shape, function, biological mechanism, and control mechanism of biological systems on the basis of robots, with the advantages of emotional interaction.
On the other hand, as for the biophilic nature of human thinking \cite{grinde2009biophilia}, bionic robots have the potential to evoke human emotions and empathetic responses \cite{daas2014toward}, interact directly with people \cite{bradwell2021morphology}, and improve human-robot interaction and user experience \cite{ghafurian2020design}.
%, a bionic robot is proposed to improve human-robot interaction and user experience through the design and evaluation of 11 emotional expressions of the animal-like robot Miro. 
However, whether the bionic form of the quadruped robot will bring good emotional support to BLV people has not been deeply explored.

To verify the functional and emotional benefits and drawbacks of the quadruped guide robot, we use the wheeled robot, another major form of mobile robot, as a comparing baseline. 
Functionally, previous studies provided evidence that the wheeled robots can provide grounded kinesthetic feedback, and have been applied to guide BLV people by researchers \cite{tobita2018structure, guerreiro2019cabot, capi2012development}. 
However, the wheeled robot lacks some terrain accessibility because of the limitation of the chassis. 
Regarding emotional assistance, wheeled robots do not have a bionic form and may be unable to provide emotional support.
%Wheeled robots can be used as baseline to verify the functional and emotional advantages of quadruped robots through comparative experiments
Therefore, we conducted this in-depth study comparing the quadruped robot and wheeled robot to further explore the implications for future guide robot design.% and expected to get the following answers:
% \begin{itemize}
%    \item Will the benefits of quadruped robot lead to better guidance for the real blind?
%    \item Will the bionic form of the quadruped robot bring more trust and satisfaction to real blind people?
%    \item Will the zoomorphic robots produce empathic responses to the blind?
% \end{itemize}

 The contributions of this paper are summarized as follows:
\begin{itemize}
    \item Functional comparison through empirical experiments on blind guidance comparing quadruped and wheeled robots.
    \item Preliminary exploration of potential causes that affects the user experience of quadruped robots on blind guidance in comparison to wheeled robots.
    % \item Exploration of current important factors that affect the user experience of quadruped robots on blind guidance in comparison to wheeled robots.
    \item Emotional needs from BLV user's perspective around bionic forms, especially in the context of guiding.
    \item Inspiration for the design of the future guide robot to improve the guiding experience while maintaining functionality.
    %bionic gait control in line with the habits of guide dogs, grounding posture conduction, and multi-modal emotional interaction.
\end{itemize}

\section{METHODOLOGY}

To explore the guiding experience differences between quadruped and wheeled robots, we conducted a within-subjects experiment including two stages: the laboratory stage and field stage, which are conducted in indoor and outdoor environments. 
We used the Wizard-of-Oz method to test both two forms, no matter the indoor tasks or outdoor tasks. And the study was conducted with the approval of our University’s Institutional Review Board (IRB).

\subsection{Research Questions} 
% We formulate the following hypotheses inspired by the questions we listed in part I. 

\textbf{RQ1:} Functionally, how many differences in guiding experience between quadruped robots and wheeled robots, and what are the potential causes?
% Under indoor conditions, people report better guidance and experience (lower workload, higher usability, trust, and satisfaction) when guided by the wheeled robot.

\textbf{RQ2:} Emotionally, what kind of emotional experience can the bionic forms of quadruped robots bring in guiding?
% Under outdoor conditions, people report better guidance and experience (lower workload, higher usability, trust, and satisfaction) when guided by the quadruped robot.

\textbf{RQ3:} What are the special insights of designing a quadruped guide robot from the perspective of BLV users?
% Blind people prefer bionic robots and empathize with them as much as discerning people do.  

\subsection{Participants}
23 participants (7 females, 16 males, 28.870 $\pm$ 6.608 years old) were invited to participate in the indoor and outdoor walking experiment (14 indoor participants, nine outdoor participants), including eight sighted participants, expected to simulate the state of suddenly acquired blindness \cite{slade2021multimodal}, and 15 blind participants. We collected basic information about the participants through a questionnaire. We screened them according to two criteria: ignorance of the experimental route and the capabilities of the two forms of robots (such as operation methods, etc.) (TABLE \ref{participant}). 

\begin{table}[b]
\caption{Participants' Demographic}
\label{participant}
\centering
\resizebox{\columnwidth}{30mm}{
\begin{tabular}{cccccccc}
\hline
    % \parbox[c][8mm]{10mm}{\centering Experiment} 
    
    & No.                  & Gender               & Age                  &  VI degree          & O\&M & Travel /week & Aids  \\ 
\hline
\multirow{5}*{INDOOR}          & P1                   & F                    & 32                   & Low vision                                            & \checkmark                              & \textgreater 5 times            & Guide dog                     \\
                                 & P2                   & M                    & 28                   & Low vision                                            & ×                              & 2-3 times                    & \textbackslash{}              \\
                                 & P3                   & F                    & 18                   & Low vision                                            & \checkmark                              & Uncertain                    & \textbackslash{}              \\
                                 & P5                   & F                    & 26                   & Blind                                       & \checkmark                              & 7 times                      & White Cane                    \\
                                 & P6                   & M                    & 34                   & Blind                                       & \checkmark                              & 4 times                      & Guide dog                     \\ 
\hline
\multirow{9}*{OUTDOOR}         & P7                   & M                    & 32                   & Low vision                                            & ×                              & 1 times                      & White Cane                    \\
                                 & P8                   & F                    & 31                   & Low vision                                            & ×                              & Hardly ever                  & White Cane                    \\
                                 & P9                   & M                    & 35                   & Low vision                                            & ×                              & 7 times                      & White Cane                    \\
                                 & P10                  & M                    & 30                   & Blind                                       & ×                              & 4 times                      & White Cane                    \\
                                 & P11                  & M                    & 34                   & Low vision                                            & ×                              & 3 times                      & White Cane                    \\
                                 & P12                  & M                    & 23                   & Blind                                       & \checkmark                              & 5 times                      & White Cane                    \\
                                 & P13                  & M                    & 49                   & Low vision & ×                              & 5-6 times                    & White Cane                    \\
                                 & P14                  & M                    & 35                   & Blind                                       & ×                              & 4-5 times                    & White Cane                    \\
                                 & P15                  & M                    & 32                   & Low vision                                            & ×                              & 2-3 times                    & \textbackslash{}              \\ 
\hline
\multirow{8}*{INDOOR}          & P16                  & F                    & 22                   & Sighted*                                    &                                &                              &                               \\
                                 & P17                  & M                    & 22                   & Sighted*                                    &                                &                              &                               \\
                                 & P18                  & M                    & 24                   & Sighted*                                    &                                &                              &                               \\
                                 & P19                  & M                    & 22                   & Sighted*                                    &                                &                              &                               \\
                                 & P20                  & F                    & 24                   & Sighted*                                    &                                &                              &                               \\
                                 & P21                  & M                    & 26                   & Sighted*                                    &                                &                              &                               \\
                                 & P22                  & F                    & 26                   & Sighted*                                    &                                &                              &                               \\
                                 & P23                  & M                    & 26                   & Sighted*                                    &                                &                              &                               \\     
\hline
\end{tabular}
}
(*: simulating suddenly acquired blindness)
\end{table}

\subsection{Robots and Control System used}

\begin{figure}[ht]
    \centering
    \includegraphics[width=0.85\columnwidth]{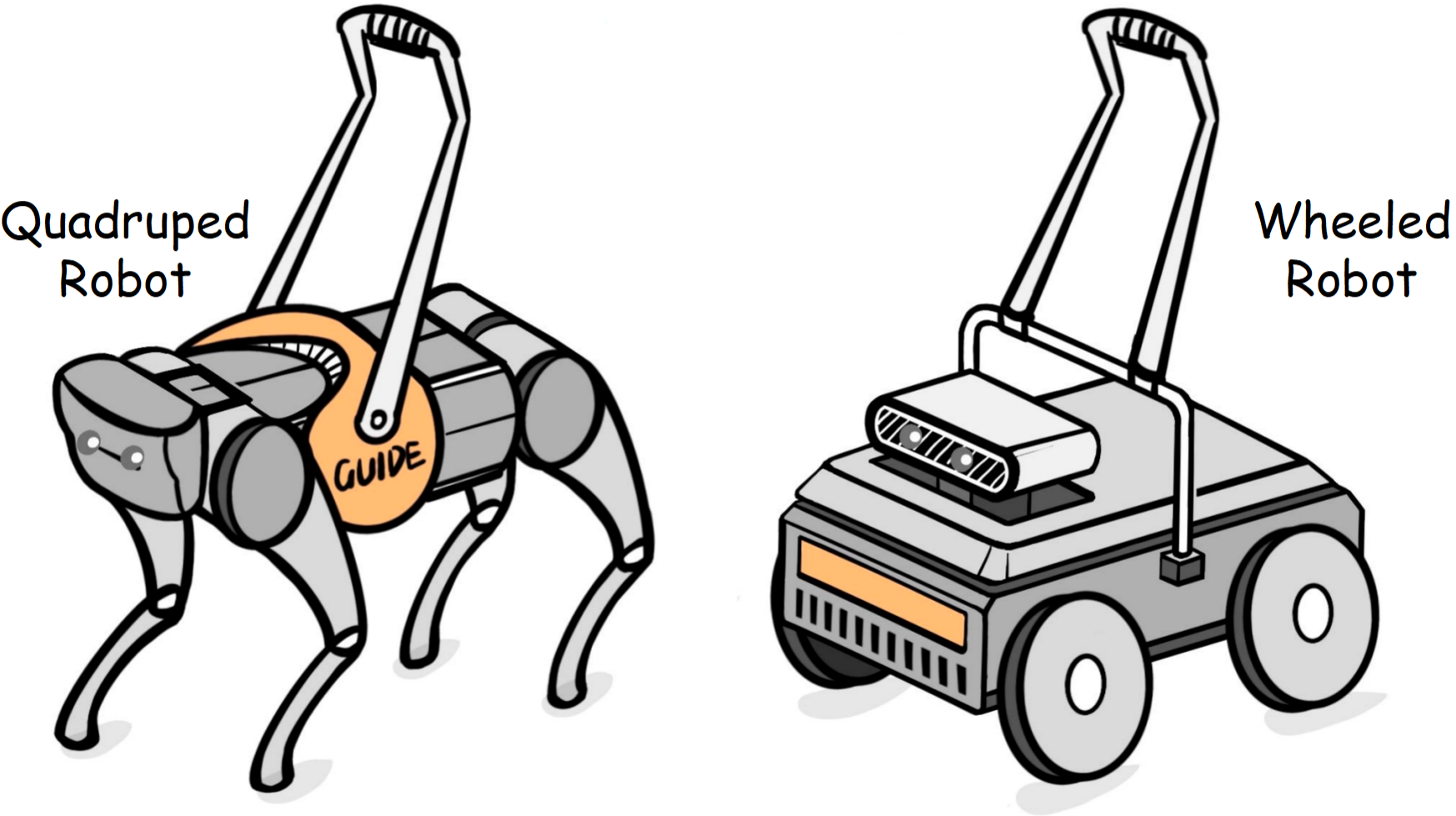}
    \caption{Quadruped robot and wheeled robot}
    \label{fig:robot1}
\end{figure}

\textbf{Quadruped Robots:} 
The quadruped robots used in the experiment are both in bionic forms (Fig.~\ref{fig:robot1}), which is a mechanical structure with 12 degrees of freedom realized by 12 high-performance servo motors. They both inherit the open source force dynamic equilibrium of MIT Cheetah \cite{cheetah2019, katz2018low}, with good multi-terrain adaptability. Basically, they can realize the necessary movements in navigation, such as forward and backward, left and right movement, turning in place, up and down slopes, etc. Leg movements are limited to various gaits. The indoor experiments used the Cyberdog from Xiaomi, 771×355×400 mm (L×B×H), while the outdoor experiments used the A1 from Unitree, 650×310×600 mm (L×B×H).

Additionally, guide dog users routinely use a guide harness, which is divided into a harness (adapted to guide dogs’ bust size of 640 to 940 mm) and a tie rod (450 mm), to perceive the guidance of the guide dog better. Therefore, the quadruped robots in this experiment are equipped with an entire guide harness like guide dogs.

\textbf{Wheeled Robots:}
As the comparison group for the quadruped robots, the size of the wheeled robot used in the indoor experiment is designed to be the same as Cyberdog, 771×355×400 mm (L×B×H), and the outdoor experiment selected the Guiding Car \cite{zhang2023follower} with the size of a guide dog as a reference to better accommodate the outdoor complicated environment. In addition, the two-wheeled robots are respectively equipped with a tie rod of guide harness and an adjustable handrail by rigid connection to achieve a similar guide experience to quadruped robots.

\textbf{Control \& Operation:}
Considering the complicated environment and technology limitations, we unified the remote control system to realize the Wizard-of-Oz method. The controller was a 360° rocker, transmitting data such as moving direction, speed, and emergency brakes to the robot wirelessly. It was tested that the operation delay of the remote control system was less than 100 ms, enough to meet the needs. During the indoor experiment, the operator stood in the non-test area and operated the robot to follow the ground route. In the outdoor experiment, the operator was always located 3m behind the participant and stopped operating for a short time in case of an emergency.
To eliminate the influence of the operator's subjective factors, in each experiment, all instructions to two kinds of robots were made by one operator, and she accepted operational training before the experiment to ensure robots guide participants fluently. 
%HC-04 Bluetooth module was used to transfer data wirelessly. The controller was a laptop, and the input was acquired through the keyboard by an operator. In addition to the basic forward speed and steering control, there were keys set for emergency brakes. After testing, the operation delay of the remote control system was less than 100 ms, which was enough to meet the needs. About the operation of robots, we considered that the balance between subjective factors and security is a topic worth discussing. Considering the specificity of the group and the complexity of the environment, during the experiment, all instructions to two kinds of robots are made by one operator and she accepted operational training before the experiment to ensure robots guide for participants fluently. 

\subsection{Experiment Environment}
\begin{figure}[ht]
    \centering
    \includegraphics[width=0.9\columnwidth]{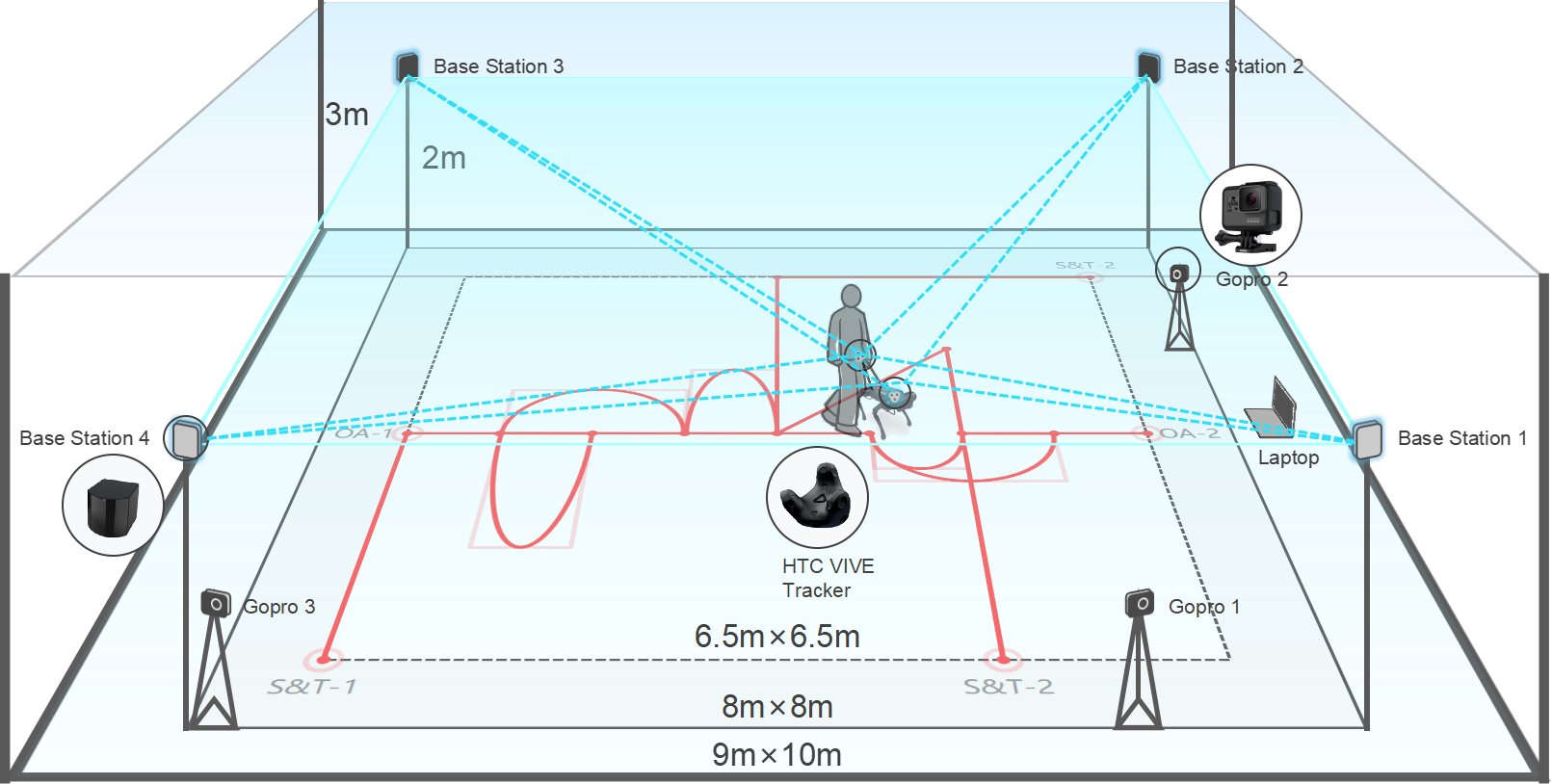}
    \caption{Schematic diagram of indoor scene layout}
    \label{fig:control_scene}
\end{figure}

\textbf{Indoor environment:}
We carved out an indoor experimental area of 8×8 m in the center of a 9×10 m lecture hall, with a 6.5×6.5 m area as the walking task area (Fig.~\ref{fig:control_scene}). The base station of HTC VIVE was installed in the four corners of the experimental area (8×8 m), 2m from the ground, to cover the entire area. Then, by tying two Trackers to the participant's waist and the robot's body, the Tracker could receive the infrared light signal emitted by the base station and give feedback. Through HTC Vive Tracker, we can obtain real-time position coordinates between participants and robots.

\begin{figure}[b]
    \centering
    \includegraphics[width=0.95\columnwidth]{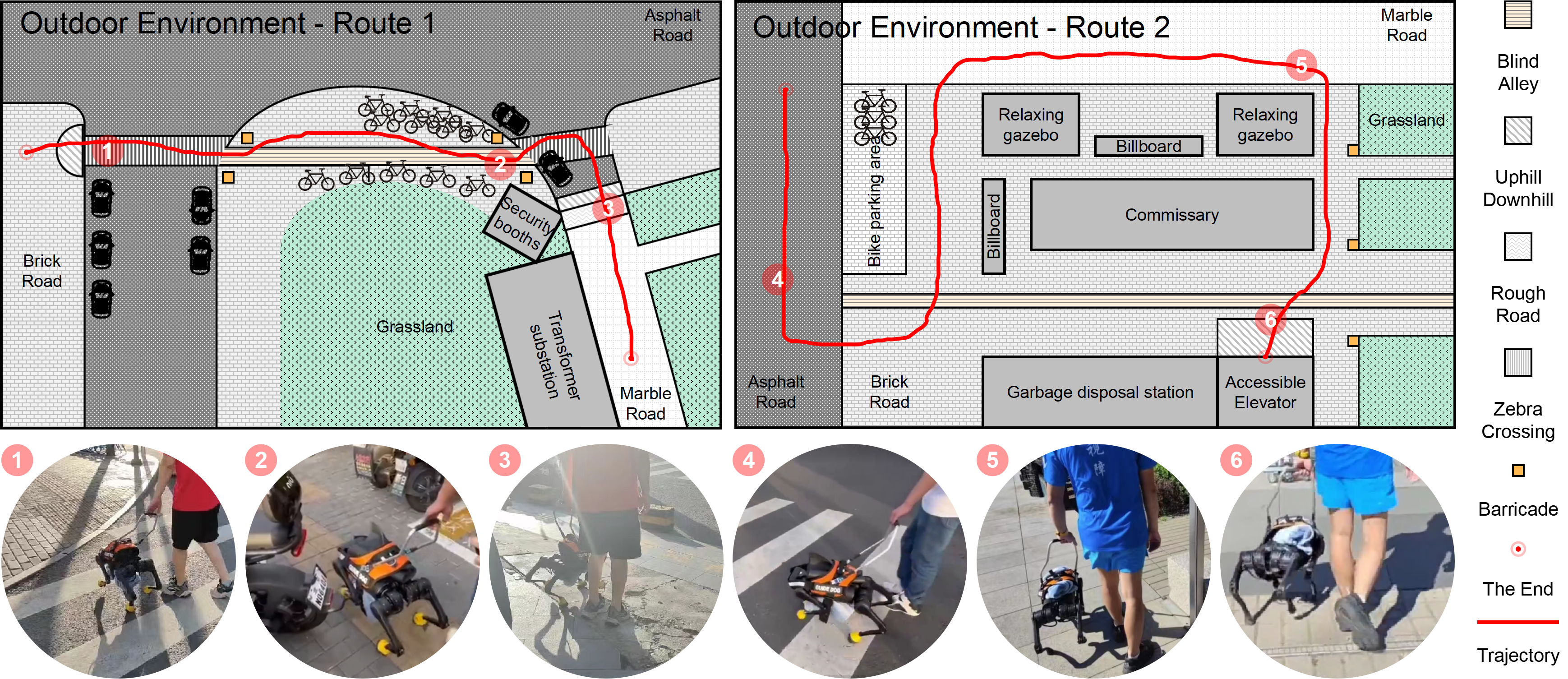}
    \caption{The outdoor experiment setup and sample trajectory.}
    \label{fig:outdoor_route}
\end{figure}

\textbf{Outdoor environment:}
To explore the performance of two forms of guide robots in real daily life, we asked the participants for some desired destinations that they have difficulties reaching alone nearby their neighborhood in advance. And two routes were planned first. %Considering the convenience of the blind and their daily walking situation, we found two areas to conduct our outdoor experiment. 
After trying to simulate the real travel conditions of the BLV user, %To control the effects of different ground conditions, 
we then made some adjustments to ensure that the two walking routes both include brick surface, uneven brick surface, tarmac, narrow road, obstacles, and uphill/downhill. These grounding conditions include the majority of areas where the BLV often visits. We did these to allow the BLV to experience as many different chassis of robots interacting with different road conditions as possible while ensuring their safety and preventing fatigue, and thus to evaluate different chassis of robots fully. We also considered stairs, but considering the limitations of current quadruped and wheeled robots, we gave up the plan of navigating the stairs. The total length of the two routes is about 100 meters. The detailed grounding information and environment can be seen in Fig.~\ref{fig:outdoor_route}.

\subsection{Procedure}

Firstly, all participants were invited to know the detail of a consent form and sign up for it. Because of the difference in visually impaired degrees, which means some participants have feelings of light or see the shadow of stuff, an eye patch was worn before the experiment. We also paid attention to the disruption of sound from robots, which were controlled by wearing earphones to hear the same sound. But in the outdoor experiment, to simulate a real outdoor walking scene and prevent risks, we didn't ask participants to wear eye-patch and earphones anymore. Then, due to challenges related to using the guiding harness, we set up a session to train the use of it. %During this period of time, the function of the guiding harness was introduced, and they could experience how to own directions information anytime until they reported they handled the device.

\begin{figure}[b]
    \centering
    \includegraphics[width=\columnwidth]{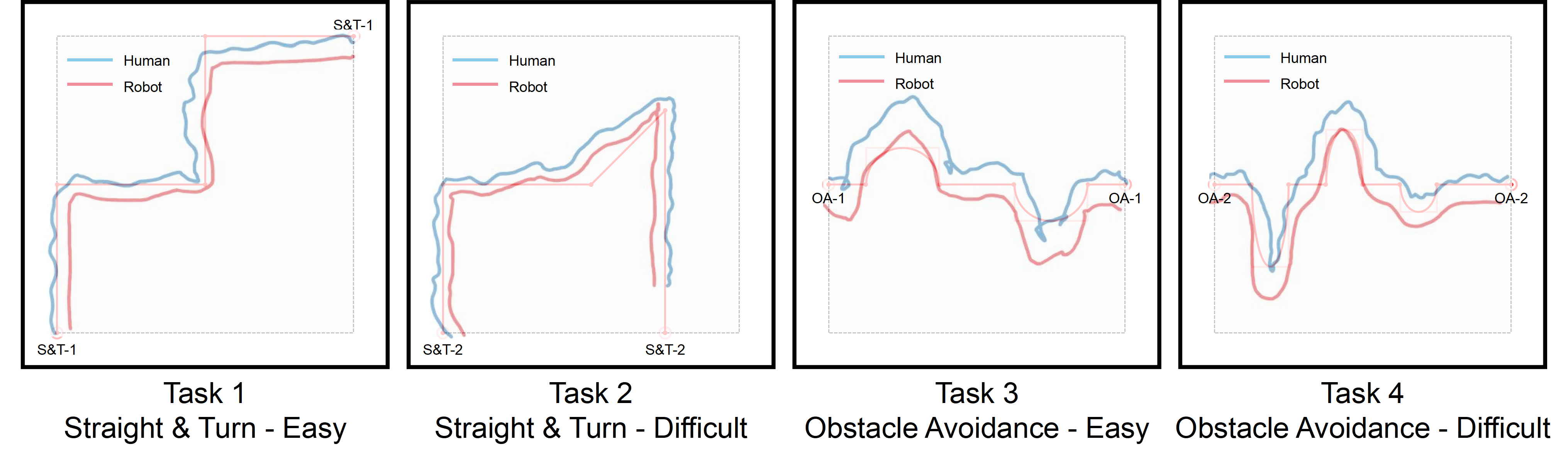}
    \caption{Demonstration of indoor navigation tasks.}
    \label{fig:indoor_task}
\end{figure}

After preparation, participants were invited to finish walking tasks. We designed an indoor study to have four trials for each chassis. Trials combining different tasks and difficulties were presented for each participant by reverse balance method to decrease sequential effects. 
For the difficulty, we used easy and difficult settings. And for the task, we used the straight \& Turn (S\&T) and Obstacle Avoidance (OA) settings. About the definition of the difficulty of tasks, we take rotation angle and the number of obstacles into consideration. For easy S\&T tasks, we arrange the robot to lead the participants to make three 90-degree turns. For difficult S\&T tasks, robots were set to turn at different angles(45 degrees, 90 degrees, and 135 degrees) \cite{liu2021tactile, yang2021lightguide, zhang2023follower}. Two identical chairs heading in different directions were placed in the laboratory for easy OA tasks. For difficult OA tasks, three identical chairs were placed, and participants had to avoid the obstacles more frequently in a short distance, and the arc length of avoidance was longer. (See Fig.~\ref{fig:indoor_task})
During the outdoor experiment, they also went straight and turned, including right angle turn, and larger and smaller angles. Avoiding obstacles also happens naturally, such as pedestrians, bicycles and wire poles. %Besides, we set the route through a variety of road surfaces and road conditions. The detailed grounding information can be seen in Fig.~\ref{fig:outdoor_route}.

After each experiment, the participants were invited to rate each robot's usability, workload, satisfaction, and trust on relevant scales (details are in part \textit{F}). Finally, the participants were invited to accept a semi-structured interview about their general experience with robots. Such questions as \textit{"what do you think of the advantages of quadruped and wheeled robots?"} and \textit{"how do you feel when you turn or avoid obstacles with quadruped robot and wheeled robot?"}

\subsection{Measures}
About the questionnaire, the subjective valuables were categorized into the below two aspects: subjective assessment of robots' use (usability, workload) and personal feeling (trust and satisfaction). All questionnaires have passed the reliability test, and their Cronbach's alpha is acceptable.

% \textbf{Workload:}
% The National Aeronautics and Space Administration Task Load Index (NASA-TLX) scales were used for assessing the ease of use\cite{hart2006nasa}. This scale includes six items: mental demand, physical demand, temporal demand, performance, effort, and frustration. All participants were asked to rate from 0 to 20.
\begin{enumerate}
    \item \textbf{Workload (NASA-TLX):} The National Aeronautics and Space Administration Task Load Index (NASA-TLX) scales were used for assessing the ease of use\cite{hart2006nasa}. This scale includes six items: mental demand, physical demand, temporal demand, performance, effort, and frustration. All participants were asked to rate from 0 to 20.

    \item \textbf{Usability (SUS):} In 1996, the System Usability Scale (SUS) was created by Brooke et al. \cite{brooke1996sus}. It assessed usability via a five-point Likert scale ranging from 1, completely disagree, to 5, completely agree, replacing all the words "system" with "robot".

    \item \textbf{Trust:} The questionnaire was adapted from the seven-point trust in automation scale by Jian et al. \cite{jian2000foundations}.

    \item \textbf{Satisfaction (QUEST):} The Quebec User Evaluation of Satisfaction with Assistive Technology (QUEST) Scale by Demers et al.~\cite{demers1996development} measured the satisfaction level. 1-5 scores were rated as the level of satisfaction.
\end{enumerate}

\section{FINDINGS}

\subsection{RQ1: Functionally, how many differences in guiding experience between quadruped robots and wheeled robots, and what are the potential causes?}
\subsubsection{\textbf{Functional differences}}
%%首先陈述结果
Before the experiment, the most prominent strengths of the quadruped robot, performing well in complex terrain, were expected to remain consistent in the field of guiding. 
Conversely, from our BLV participants' perspective, \textbf{wheeled robots received significantly higher user evaluation than quadruped robots}.
%could maintain stability in an indoor guiding environment, while quadruped robots do not obviously have outstanding dominance. 
The outdoor environment did not change the evaluation trend but made the differences more pronounced. The details are as follows.

\begin{figure}[b]
    \centering
    \includegraphics[width=0.95\columnwidth]{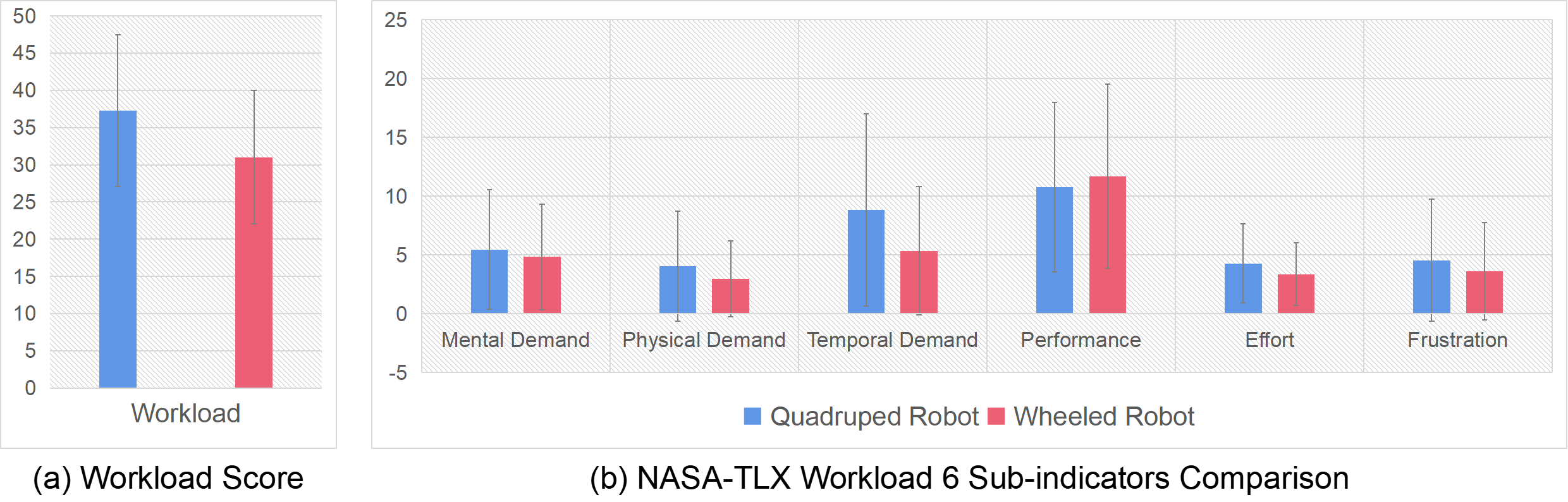}
    \caption{Results of the indoor experiment.}
    \label{fig:indoor_result}
\end{figure}

%%具体的数据结果
We performed matched samples t-tests to determine the statistical significance according to questionnaire scores. The \textbf{(1) NASA-TLX} score showed the perceived workload of participants showed significant differences between the two forms of robots. The wheeled robot showed a lower workload than the quadruped robot (t=-2.598; p\textless0.05) (see figure \ref{fig:indoor_result}). When guided by the quadruped robot, they felt more temporal demand than the wheeled robot (t=-2.136; p=0.05). There were trends on other dimensions, but no statistical significance was found.
% But we didn't determine a statistical significance between the two forms on usability, satisfaction, and trust. Notably, there were trends (see Fig.~\ref{fig:indoor_result}). 

% No matter what aspects, the wheeled robot always showed higher average scores than the quadruped robot. 

For robot usability, we conducted a common analysis method on the \textbf{(2) SUS} results to reveal the information behind trends.
We transformed the SUS grades into different levels (A to F) according to the SUS grading range of fractional curve, and the level D \& F can be described as "bad" and "terrible" in the description of SUS scale \cite{bangor2008empirical, bangor2009determining}. The final results showed that the quadruped robot was the higher one evaluated as the D \& F (35.7\% \& 21.4\%) followed by the wheeled robot (21.4\% \& 21.4\%). We also analyzed the difference in the sub-scales of the SUS scale, effectiveness \& learning ability, use efficiency \& usability, and satisfaction. The majority of participants (78.57\%) gave higher or equal scores to the wheeled robot no matter what sub-scales they rated. For satisfaction and trust, the wheeled robot was better or equal to the quadruped robot (64.29\%). And the satisfaction result is similar to the result of the trust. The majority of participants trusted the wheeled robot more (64.29\%).

\begin{figure}[b]
    \centering
    \includegraphics[width=0.95\columnwidth]{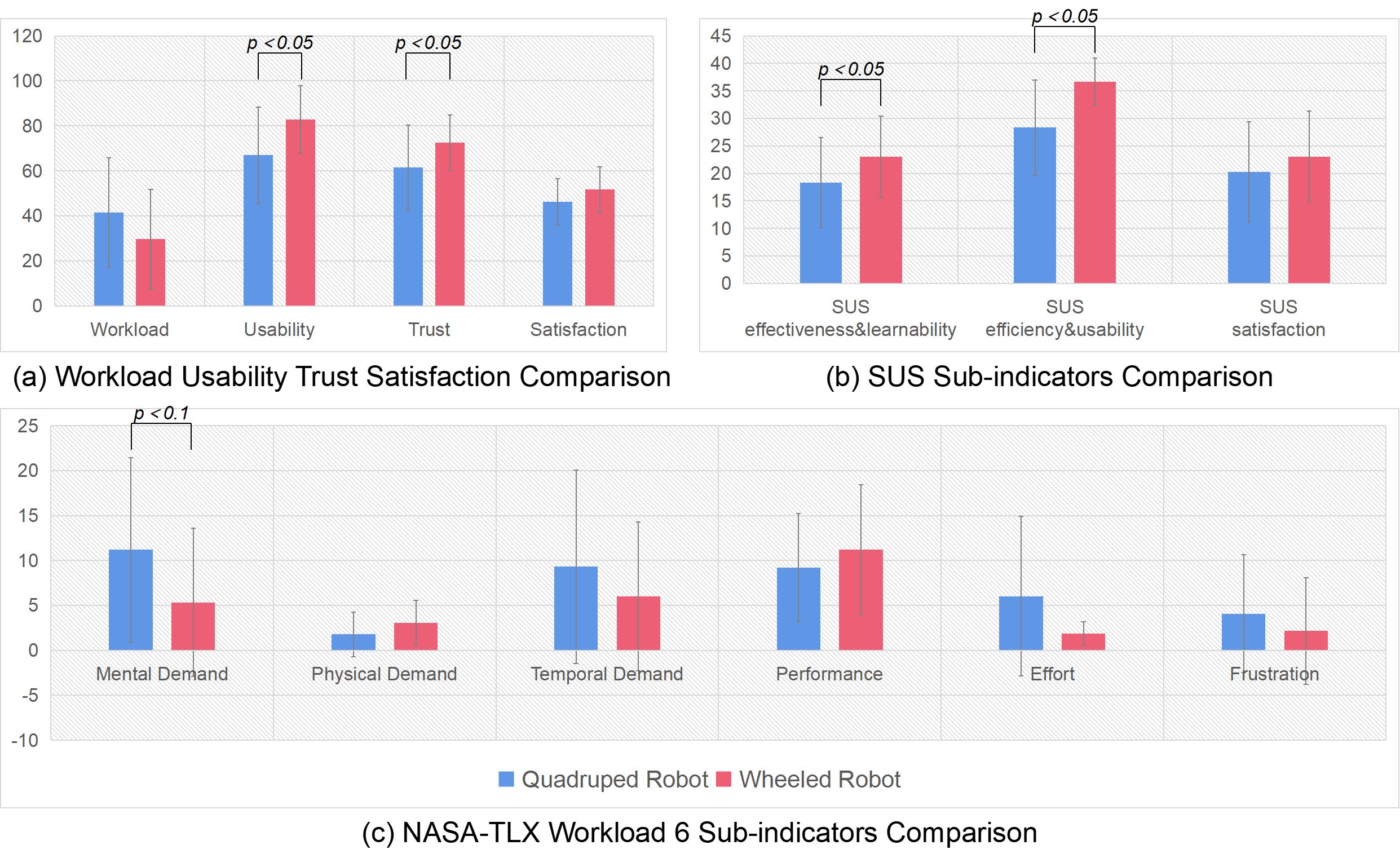}
    \caption{Results of the outdoor experiment.}
    \label{fig:outdoor_result}
\end{figure}

In the outdoor environment, the statistical results showed that the wheeled robot gave participants significantly more usability (t=-2.610; p\textless0.05) according to \textbf{(2) SUS} and more \textbf{(3) Trust} (t=-2.325; p\textless0.05). However, the workload from \textbf{(1) NASA-TLX} and satisfaction from \textbf{(4) QUEST} did not show a significant difference between the two forms. Contrary to the expectancy, the quadruped robot did not perform better in the outside experiment. In detail, participants gave significantly higher scores to the wheeled robot on effectiveness \& learning ability (t=2.341; p\textless0.05) and use efficiency \& usability (t=3.123; p\textless0.05). Although the total workload score wasn't significantly different in the outdoor experiment, one of the six sub-scale scores, mental demand, was close to significant (t=2.207; p=0.058). The mental load of using the quadruped robot is bigger than the wheeled robot. (see Fig.~\ref{fig:outdoor_result})
We also explored whether the differences existed between different groups (the suddenly acquired blindness, the blind walking with a guide dog, and the normal blind). Even though there was no statistical significance in between, the above trend is consistent.

\subsubsection{\textbf{Potential Causes}}
Based on the interviews, we analyzed why the quadruped robot did not perform better in the guiding experiment. 

%%步态问题
Firstly, \underline{\textbf{moving gait}} could be key in affecting participants' experience. The typical gaits of quadruped robots include walking, trotting, flying trot, bounding, galloping, free gait, etc\cite{gehring2014towards,di2018dynamic,fukuoka2009dynamic}. Moving gait is the most frequent motion in guiding. Since the Moving gait is a sequential collection of the support states and shifting states of its legs over time to keep balance, its motion is less smooth than that of the wheeled robot. 

During the experiment, the participants sometimes felt anxious about the behavior of quadruped robot, just like P16 said, \textit{"When the dog was walking, I was particularly afraid that it stepped on my foot, stepping on me should be more painful than if the wheel rolled over me, anyway, I was very anxious."} This provides some insight into our interpretation of the participants' lower trust in the quadruped robot. The inharmonious movement way itself and the feeling of risk generated by this way of going forward are both possible reasons for decreasing the trust scores of the quadruped robot. Furthermore, the vibrations that can not provide useful information from the ground during the walking process also affected the perception of the BLV people, even though it's tiny and imperceptible from sighted people's perspective. 

The outdoor results showed that the significantly higher time demands in the indoor experiment disappeared and were replaced by a higher mental load, indicating there was no feeling of a rush. Although the outdoor environment added a variety of factors to focus on when using both robots, such as urban noise, which made the motor sounds seem more acceptable, only the mental load of the quadruped robot substantially improved. P14 said \textit{"It seemed to be hesitating, which made me lose a little bit of confidence in it."} These additional thinking processes and speculation may increase the load and undermine trust. Although the same operator used the same operating criteria, the experimenter's subjective influence must be considered. \cite{wang2018my} mentioned that sophisticated sensing equipment is often the initial impression of a machine-like robot, making people naturally trust the robot. Researchers also found that the degrees of trust is most pronounced at the start of the interaction. Instead, the zoomorphic cognition process, a Type 1 automatic process \cite{spatola2022cognitive}, possibly makes participants feel "it is hesitating." Thus fear and confusion that emerged during this experiment may be temporary. It remains to be confirmed how BLV people feel in long-term use.  

Besides, the moving gait also affects feedback. Transverse degrees of freedom of the hip joints is essential if robots are expected to walk as flexibly as real animals. Lateral degrees of freedom introduces additional rocking motions to the robot. But P22 mentioned \textit{The dog wiggling from side to side may provide extra information to mislead me.} Moreover, the interview after the outdoor experiment concluded that the quadruped robots could barely transmit ground information to the users. At the same time, wheeled robots are more sensitive in informing participants of road surface information, such as roughness, up and down hills, etc. Sometimes we tend to consider feedback necessary, but real preferences for this feedback vary. Some participants think \textit{"it's great, I adjust the walking speed with the help of the information'}, but at the same time, some consider that \textit{"it doesn't matter"} or even \textit{"I don't like the feeling, the surface is not what I need to know, just let me pass."} Thus, it can be studied further about what kind of and how much information should be given for the BLV.

%%噪音问题
Secondly, \underline{\textbf{walking noise}} generated by the robot could be a huge problem disrupting the walking process. The leg form of a quadruped robot makes them withstand impact from different directions. Thus, the impact on the ground during walking recoils the motors at the joints, generating loud and unavoidable noises. The feet of real animals are usually irregular, with structures such as claws and fleshy pads, and always have a strong grip and remain relatively silent. However, the truly bionic feet have not been applied to most quadruped robots~\cite{tian2020bionic}.

\begin{figure}[ht]
    \centering
    \includegraphics[width=\columnwidth]{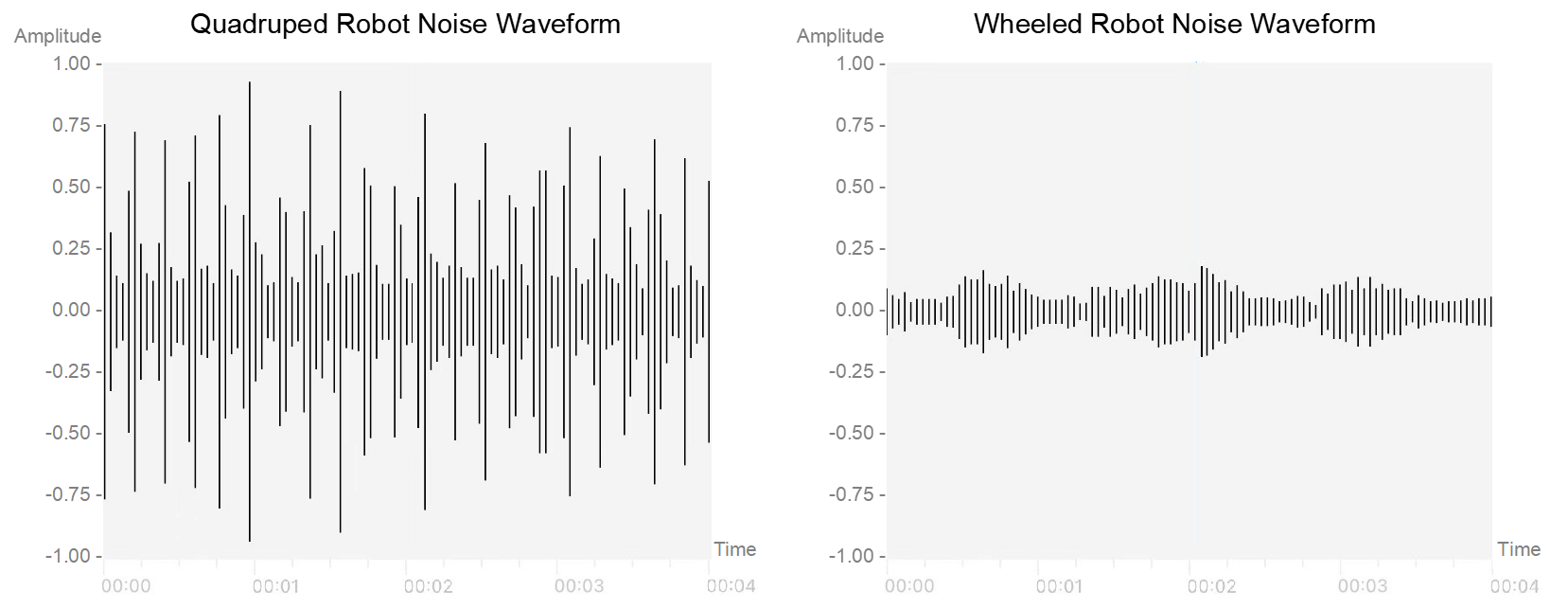}
    \caption{Illustration of the operating noise waveform of the quadruped (left) and wheeled (right) robot.}
    \label{fig:noise}
\end{figure}

The sound intensity of both wheeled and quadruped robots in operation was measured. The average and maximum noise for the quadruped guide robot in the indoor experiment environment were 80 dBA and 82 dBA, while the average and maximum noise for the wheeled robot was 65 dBA and 70 dBA. The indoor environmental noise was 36 dBA. The pattern of the sound waveform can be seen in Fig.~\ref{fig:noise}. The wheeled guide robot sounded quieter and more subdued.

% Before the experiment, we also mentioned the problem but were unsure of the degree of influence. 

According to the interview, some participants expressed that \textit{"The dog is noisy"}, though they all wore headsets. Hearing is a significant sensory resource for the BLV, meaning this repeated and meaningless sound will likely disrupt their access to outside feedback, especially in a relatively quiet environment. Also, the problem posed by the sound affected their social life, just like P6 said, \textit{"On the one hand, I always have to stand it (the sound). On the other hand, for example, what do we do when we attend a meeting or go out in the process of a meeting? A real dog can be very quiet (...), but this robot is loud, which makes me feel embarrassed."} 
%%That inspired us to take more broaden possible occasions of use into consideration, guiding experience happens both in the process of guiding and before and after use. And these occasions are not single because the BLV users sometimes share similar social life occasions with most people. Quadruped robot does well in dealing with complex terrain conditions, but social life is similarly complex to think about it.

\subsection{RQ2: Emotionally, what kind of emotional experience can the bionic quadruped robots bring in guiding?}

Previous studies showed that older people preferred realistic, familiar, bionic designs over mechanomorphic, and such designs increased the incidence of emotional responses and interactions\cite{bradwell2021morphology}. 
% Also, bionic robots can play games with children with autism and support the development of joint attention\cite{szymona2021robot}.
However, we did not know how robots in dog form would affect the BLV users. Thus, we studied whether the group could also empathize with quadruped robots. 
One observation was that \textbf{most participants can neither feel quadruped robots were friendly nor could bring them a range of positive emotions}, such as companionship and cordial feeling. Notably, one of the themes repeatedly emphasized was practicality, like P10 said, \textit{"The appearance is not important as long as it's portable. It is the guiding capability that matters."} According to the description from guide dog users, despite their passion for guide dogs, they have no preference for the bionic form. 
% Conversely, P6 mentioned \textit{"In fact, I didn't like the dog at first, and I even resisted a little, if there would be a tool to replace it in the future, that must be good, and if I really like dogs, I will get a pet at home, why do they have to do this (guiding)."} ,
Whereas, there were still two participants( P17 \& P21) who expressed that the robotic dog was affectionate, which they described as if they were walking a dog, yet they were both sighted people. Similarly, previous research had organized focus groups to learn about BLV people. When the participants expressed their perception of a guide robot, some BLV people imagined that they might want a dog-like robot \cite{bonani2018my}. However, when the BLV people in this experiment used the dog-like robot, they realized safety and comfort were the most important guiding aspect. 
%%This provides us with the inspiration that though the combination of the "animal form" and "walking" follow the popular trend of friendly human-machine interaction and emotion design\cite{biswal2021development}, we still explore a better time to talk about the topic because we still can't find a perfect solution to the great risks during guiding. And whether the widely discussed question, valley of terror effect, can exist in the BLV also deserve to study in the future work. That determines the similarity between domestic bionic robots and real animals.

\subsection{RQ3: What are the special insights of designing a quadruped guide robot from the perspective of the BLV?}
The above finding revealed the potential improvement direction for quadruped guide robots. The underlying design logic of quadruped robots gives them the ability to make greater breakthroughs in finishing complex tasks, such tasks as getting on the steps\cite{winkler2015planning}, trotting over obstacles\cite{barasuol2013reactive}, dynamic grasping\cite{zimmermann2021go}, and so on. Therefore, we hope the following insights could be incorporated into the future iteration of quadruped guide robots to improve the walking experience while maintaining functionality.

Developing harmonic and gentle \underline{\textbf{moving gaits}} could decrease unnecessary feedback and enhance the sense of certainty. Naturally, anxiety from being stepped on could relieve it to a larger extent. However, a meaningful topic is what kind of and how much information is appropriate for BLV users, and will various feedback lead to being confused or feeling safe? Moreover, adding professional actions of guide dogs, such as stopping on the front paws before going up the steps, will also greatly help the BLV users.

Besides, participants repeatedly mentioned the \underline{\textbf{noise problem}} inspired us to consider broad possible occasions of use as BLV users sometimes share similar social life with most sighted people. Quadruped robot does well in complex terrain conditions, but social life is similarly complex to think over. The application of irregular and variable shapes or flexible materials can be one solution to reduce the walking noise of quadruped robots. 

Simultaneously, the findings on BLV users' \underline{\textbf{emotions}} toward quadruped guide robots give an inspiration that even though the combination of "dog-like" and "walking" is associated with the implementation of quadruped robots as guide dogs, much efforts on the more realistic biological-structure simulation is still needed to give BLV users a similar sense of trust and satisfaction as biological dogs.
% Simultaneously, the findings about empathetic emotion give an inspiration that though the combination of the "animal form" and "walking" follow the popular trend of friendly human-machine interaction and emotion design\cite{biswal2021development}, we still explore a better time to talk about the topic because we still can't find a perfect solution to the great risks during guiding. And whether the widely discussed question, valley of terror effect, can exist in the BLV also deserve to study in the future work. That determines the similarity between domestic bionic robots and real animals.

Other robot chassis can be considered in the future, such as \underline{\textbf{wheel-legged robots}}, which compensate for the weaknesses of each independent chassis. Now wheel-legged robot combines the strengths of both chassis like it will move in rough terrain and roll in a flat surface\cite{biswal2021development}. \cite{zhu2018stability} developed Wheel-Track-Leg Hybrid Mobile Robot with great obstacles-crossing capability, and \cite{9536060} developed a control system to customize the tail according to obstacle parameters. These breakthroughs are all beneficial for solving problems during guiding. Therefore, it would be a great choice to provide better guiding performance and experience with the help of wheel-legged robots.

%We can both amplify the functional advantages of the quadruped type and consider aspects of it that are not as good as wheeled type when guiding, such as the stability, silence, and direct feedback transfer and so on.

\section{LIMITATIONS}
The research tried to figure out the effects of quadruped and wheeled chassis, but something remains unknown in the scope of the experiments. 
Under the limitation of commercial robots in the market, the quadruped robots we chose were not developed deliberately for the BLV user. We must admit that guide robots targeting is group may make results clearer and more direct. In addition, all participants never got in touch with these robots before, so it is necessary to explore their reactions without the novelty effect of robots. In the future, long-term studies conducted in real life may be a key process to making these assistive robots a part of BLV people's lives.

\section{CONCLUSIONS}
In two studies, we conducted indoor and outdoor walking experience experiments for BLV participants using quadruped and wheeled robots. The results showed that participants preferred the wheeled robot in the indoor experiment, and this trend was reinforced by statistically significant differences in the outdoor experiment. Based on the result, we found that the unavoidable moving gait and walking noise of quadruped robots may be the obstruction during navigation. According to the discussion, we proposed some implementations for design, such as more realistic biological-structure simulation and so on. Further research is expected to study whether empathetic emotion will play a role in BLV people with the help of guide robots with enhanced functionality. Moreover, we must determine whether these findings will apply in the long-term use and larger sample size.

%Although these conclusions are preliminary and limited by small sample size and availability of devices, these could be implications for design of guide robots.

%%%%%%%%
\bibliographystyle{IEEEtran}
\bibliography{IEEEfull,root}

\end{document}